\documentclass[%
aip,
amsmath,amssymb,
reprint,%
]{revtex4-1}

\usepackage{graphicx}
\usepackage{dcolumn}
\usepackage{bm}

\usepackage[utf8]{inputenc}
\usepackage[T1]{fontenc}
\usepackage{mathptmx}
\usepackage{etoolbox}

\makeatletter
\def\@email#1#2{%
	\endgroup
	\patchcmd{\titleblock@produce}
	{\frontmatter@RRAPformat}
	{\frontmatter@RRAPformat{\produce@RRAP{*#1\href{mailto:#2}{#2}}}\frontmatter@RRAPformat}
	{}{}
}%
\makeatother

\newcommand{\NTT}{
	\affiliation{NTT Basic Research Laboratories, NTT Corporation, 3-1 Morinosato-Wakamiya, Atsugi, Kanagawa, 243-0198, Japan}
}
\newcommand{\Shizuoka}{
	\affiliation{Research Institute of Electronics, Shizuoka University, 3-5-1 Johoku, Chuo-ku, Hamamatsu, Shizuoka 432-8011, Japan}
}

\begin{document}

\preprint{AIP/123-QED}

\title{Identifying impurities in a silicon substrate by using a superconducting flux qubit}
\author{Hiraku~Toida}
\email{hiraku.toida@ntt.com}
\NTT
\author{Kosuke~Kakuyanagi}
\NTT
\author{Leonid~V.~Abdurakhimov}
\altaffiliation{Current address: IQM Quantum Computers, Keilaranta 19, Espoo, 02150, Finland}
\NTT
\author{Masahiro~Hori}
\Shizuoka
\author{Yukinori~Ono}
\Shizuoka
\author{Shiro~Saito}
\NTT

\date{\today}

\begin{abstract}
	A bismuth-doped silicon substrate was analyzed by using a magnetometer based on a superconducting flux qubit.
	The temperature dependence of the magnetization indicates that the silicon substrate contains at least two signal sources, intentionally doped bismuth spins and a spin 1/2 system with a ratio of 0.873 to 0.127.
	In combination with a conventional electron spin resonance spectrometer, a candidate origin of the spin 1/2 system was identified as a dangling bond on the silicon surface.
	In addition, the spin sensitivity of the magnetometer was also estimated to be 12 spins/$\sqrt{\mathrm{Hz}}$ by using optimized dispersive readout.
\end{abstract}

\maketitle

Superconducting qubit technology has rapidly developed as a component of quantum computers \cite{Arute2019, Kjaergaard2019} and annealing machines \cite{Hauke2020, King2024}.
One of the limitations on the performance of such machines is the quantum coherence bounded by the energy relaxation time.
In order to improve the coherence time, much effort has been spent on developing the fabrication process of qubits.
To date, research has centered on reducing losses from two-level systems at the substrate-air, metal-substrate, and metal-air interfaces, because their effects are substantial due to their high participation ratio \cite{Woods2019, Siddiqi2021}.

The energy relaxation time of superconducting qubits has surpassed 0.5 ms, with progress in advanced materials engineering \cite{Wang2022}.
However, on such time scales, the bulk loss of the substrate affects the coherence time, besides the interface losses mentioned above.
One such loss sources in silicon devices is residual impurities in the substrate \cite{Zhang2024}.
Current techniques, such as secondary ion mass spectroscopy (SIMS), can be used as a probe for these impurities. However, in order to achieve a coherence time >10 ms, their sensitivity should be improved so that they can detect small amounts of impurities ($<10^{11}$ cm$^{-3}$) \cite{Zhang2024}.

On the other hand, superconducting circuits can themselves be probes of impurity spins in semiconductors.
For example, an electron spin resonance spectrometer using a superconducting resonator with a quantum limited amplifier \cite{Bienfait2016, Probst2017a} or microwave single photon counter \cite{Albertinale2021, Wang2023} can detect electron spins with high sensitivity.
A superconducting flux qubit can also operate as a spin sensing device \cite{Toida2019, Budoyo2020, Toida2023} because of its high susceptibility to external flux.

In the study reported in this letter, a bismuth-implanted silicon substrate was analyzed using a superconducting flux qubit to evaluate the device's suitability as a detector of spins in the substrate.
The temperature dependence of the magnetization suggests that the substrate contains at least two different spin species, bismuth spin and a spin 1/2 system without a  hyperfine interaction.
By fitting the magnetization curve to theory, the ratio of bismuth spins to those of the spin 1/2 system was derived.

We also report that the spin sensitivity can be refined to 12 spins/$\sqrt{\mathrm{Hz}}$ by introducing dispersive readout instead of using a direct-current superconducting quantum interference device (dc-SQUID) or Josephson bifurcation amplifier as in the previous reports \cite{Toida2019, Budoyo2020}.

\begin{figure}[!h]
	\centering
	\includegraphics{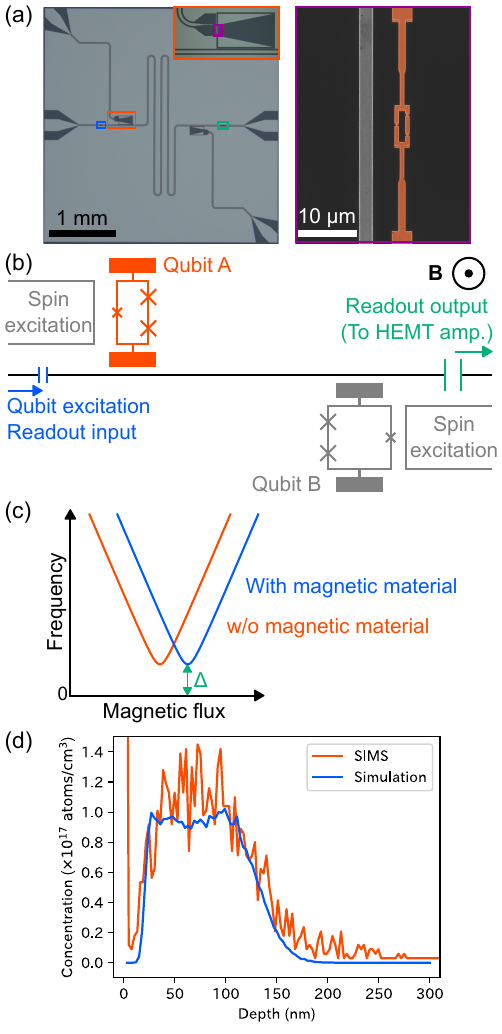}
	\caption{
		Experimental setup.
		(a) Optical and scanning electron microscope (SEM) images of the device. The blue (green) box indicates the position of the input (output) capacitor. (Inset) Magnified view of a capacitively shunted flux qubit. Further magnified image of the flux qubit loop (false-colored) is shown in the purple box.  
		(b) Schematic diagram of the qubit chip. Filled rectangles (red and gray) indicate shunting capacitors of flux qubits. In this experiment, qubit B and spin excitation lines are not used.
		(c) Flux qubit spectrum with and without magnetic material.
		(d) Measurement and simulation of the bismuth ion implantation profile. 
		The profile was measured by secondary ion mass spectroscopy.
		}
	\label{f1}
\end{figure}
A magnetometer based on a capacitively shunted superconducting flux qubit \cite{You2007, Yan2016} was used to measure the spins in a silicon substrate.
The flux qubit was fabricated directly on a bismuth doped silicon substrate [Fig. \ref{f1} (a)].
The size of the qubit loop in qubit A is 1$\times$5$\mu$m [Fig. \ref{f1} (b)].
This determines the detection area.
A superconducting resonator capacitively coupled to a flux qubit was fabricated on the same chip to read out the quantum state of the flux qubit.
To control the operating point of the flux qubit and the polarization of the impurity spins in the silicon substrate, a magnetic field ($\sim0.2$ mT) was applied perpendicular to the substrate surface [Fig. \ref{f1} (b)].
Experiments were carried out in a dilution refrigerator at temperatures from 30 to 200 mK.

A typical spectrum of a flux qubit is shown in Fig. \ref{f1} (c).
The spectrum is characterized by two parameters: persistent current $I_\mathrm{p}$ and energy gap $\Delta$. The resulting transition frequency of the flux qubit $f$ is expressed as 
\begin{equation}
	f=\sqrt{\varepsilon(\Phi)^2 + \Delta^2},
	\label{eq1}
\end{equation}
where $\varepsilon(\Phi)=2I_\mathrm{p}(\Phi-\Phi_0/2)/h$, $\Phi$ is the applied magnetic flux, $\Phi_0 = h/2e$ is the magnetic flux quanta, $h$ is the Planck constant, and $e$ is the elementary charge.
In the case of a magnetometer, $I_\mathrm{p}$ plays the most important role.
This is because a larger persistent current allows for a stronger coupling between the qubit and spins.
The flux qubit operates near an optimal flux of $\Phi= (n+ 1/2) \Phi_0$, where $n$ is an integer.
If the operation flux is away from the optimal point ($\varepsilon(\Phi)\gg\Delta$), the spectrum shows a finite slope ($\sim2I_\mathrm{p}/h$) as a function of the magnetic flux.
This slope converts the change in the magnetic flux to a change in the resonance frequency of the flux qubit.

In normal operation of a flux qubit, the flux is controlled by a small superconducting magnet attached to the sample holder.
In addition, if the magnetic material is within the range of qubit detection ($\sim \mu$m), the operating point of the qubit will be shifted due to the magnetic flux generated by the magnetic material.
In this experiment, the magnetization of impurity spins in the silicon substrate affects the operating flux of the qubit.
Thus, the qubit functions as a magnetometer for magnetic materials present in the vicinity of the qubit.

Si(100) substrates containing bismuth impurities were prepared by ion implantation with the total dose of 1$\times$10$^{12}$ cm$^{-2}$.
Figure \ref{f1} (d) shows the designed and measured implantation profiles.
After implantation, the substrate was annealed in a nitrogen atmosphere at 900 degrees Celsius for 15 minutes to activate the spins.
The activation ratio was estimated to be $\sim$50\% by conventional X-band electron spin resonance spectroscopy.
It is worth mentioning that the profile did not change significantly after annealing, in agreement with previous reports in the literature \cite{Weis2012}.

\begin{figure*}
	\centering
	\includegraphics{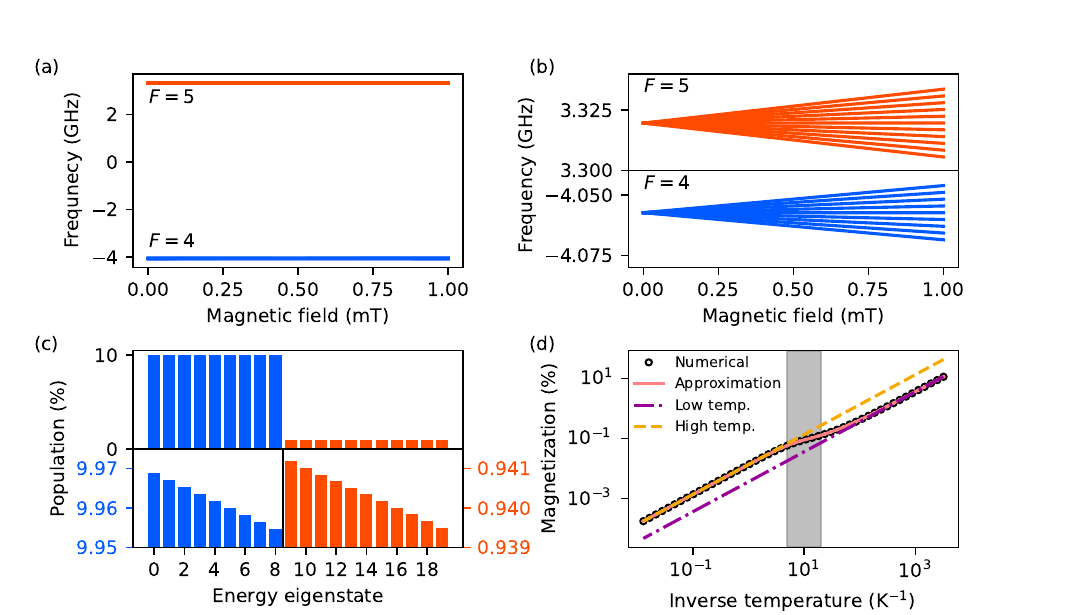}
	\caption{
		Properties of bismuth impurities in silicon substrate at low temperatures.
		(a) Energy diagram of bismuth impurities in silicon substrate as a function of magnetic field strength $|B|$.
		(b) Magnified view of the manifolds for $F=5$ (top) and $F=4$ (bottom)  in (a).
		(c) Examples of thermal population at 150 mK and 0.2 mT.
		Lower left (right) panel: Enlarged view of the $F=4$ (5) population.
		(d) Temperature dependence of normalized magnetization calculated under a magnetic field of 0.2 mT. 
		The temperature in the experiment shown in Fig. \ref{f3}(b) was swept within the shaded area.
		}
	\label{f2}
\end{figure*}
The magnetization from the bismuth impurities is calculated as follows.
First, the energy eigenvalues under a magnetic field are calculated.
The eigenvalues are then used to calculate the partition function and corresponding density matrix at finite temperature.
After that, the ensemble average of the observable $\hat{S}_z$ is calculated from the density matrix and converted into units of magnetization.

The energy eigenvalues of the bismuth impurity in a silicon substrate can be calculated using the following spin Hamiltonian:
\begin{equation}
	\hat{H} = g \mu_B \bm{B}\cdot\hat{\bm{S}} + A\hat{\bm{I}}\cdot \hat{\bm{S}}, 
	\label{eq2}
\end{equation}
where $g$ is the g-factor, $\mu_B$ is the Bohr magneton, $\bm{B}$ is the magnetic field, $\hat{\bm{S}}$ is the electron spin operator, $A$ is the isotropic hyperfine constant, and $\hat{\bm{I}}$ is the nuclear spin operator.
In particular, the calculation used $g=2.0003$ and $A/h=1475.4$ MHz \cite{Feher1959}.
Figures \ref{f2} (a) and (b) show the energy diagrams of a bismuth impurity as a function of the magnetic field strength $|B|$.
Since $S=1/2$ and $I=9/2$ for bismuth impurity, $(2S + 1)(2I + 1)=20$ energy levels are visible in total.
Depending on the total angular momentum $F = I-S$ or $I+S$, these energy levels are classified into two manifolds $F = 4$ and 5, and the corresponding numbers of energy levels, $2F+1$, are 9 and 11.

In order to evaluate the magnetization of bismuth impurities, the temperature dependence of the energy level occupancy was calculated and converted into a density matrix.
The temperature dependence at the low and high temperatures ($T\rightarrow 0, \infty$) has a simple origin as follows .
At low temperature, the intra-manifold thermal transition in $F=4$ is dominant, and the inter-manifold transition from $F=4$ to $F=5$ is negligible.
On the other hand, the inter-manifold thermal transition is dominant in the high temperature region, and $F=4$ is almost uniformly occupied.

This situation is more complicated in the intermediate temperature range $T\sim A / k_B$, where $k_B$ is the Boltzmann constant.
Since the hyperfine interaction splits the two energy manifolds with a separation of $A(I+1/2)/h\sim7.4$ GHz, $F=4$ and 5 manifolds are partially occupied at temperatures of $\sim$100 mK.
In this temperature range, crossover effects of intra-manifold and inter-manifold excitation were observed.
An example of the thermal distribution at 150 mK is shown in Fig. \ref{f2}(c).
In this case, both manifolds are partially filled.
This type of excitation thus has a temperature-dependent effect on the magnetization that can be used to distinguish different spin species.
It is important to note that this feature can be used to distinguish different spin species in the substrate.

Here, let us calculate the temperature dependence of the magnetization under a magnetic field of 0.2 mT numerically and analytically [Fig. \ref{f2} (d)].
From the spin Hamiltonian [Eq. (\ref{eq2})], an approximated form of the normalized magnetization $m$ can be derived as follows:
\begin{equation}
	\begin{split}
		m(x) &= m_L(x) + m_H(x)\\
		m_L(x) &= \frac{g\mu_BB}{A}\frac{16(I+1)+(2I+1)(2I-1)x}{6(2I+1)^2} \\
		m_H(x) &= \frac{g\mu_BB}{A}\frac{2(x-4)}{3(2I+1)} \varsigma_1\left[
		-\left(I+\frac{1}{2}\right)x+\ln\left(\frac{I+1}{I}\right)
		\right]
	\end{split}
	\label{eq3}
\end{equation}
where $x = A/k_BT$ is the reduced inverse temperature and $\varsigma_1(z)=[1+\exp(-z)]^{-1}$ is the standard sigmoid function.
In deriving Eq. (\ref{eq3}), a weak magnetic field ($g\mu_BB \ll A$) is assumed.
The magnetization can be approximated by a linear function of inverse temperature at high (low) temperature.
Only the $m_L(x)$ term remains in Eq. (\ref{eq3}) at low temperature.
In the case of $I=9/2$, the slope is 2/15.
At high temperature, $m(x)$ has a universal expression independent of $I$ and $A$, as follows:
\begin{equation}
	m(T) = \frac{g\mu_BB}{2k_BT}.
	\label{eq4}
\end{equation}
It is worth mentioning that Eq. (\ref{eq4}) is the same as the case for the spin 1/2 system without a hyperfine interaction.
The difference in the slopes (2/15 and 1/2) corresponds to the offsets of the lines for low and high temperatures in Fig. \ref{f2}(d).
A kink structure corresponding to the crossover between intra-manifold and inter-manifold excitations appears around 100 mK in this parameter range.

\begin{figure}
	\centering
	\includegraphics{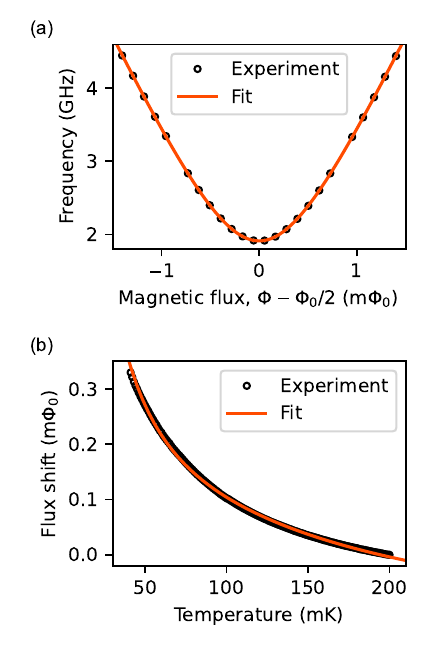}
	\caption{
		Experimental results.
		(a) Spectrum of the flux qubit. The solid line is a fitting to the data using Eq.(\ref{eq1}). 
		(b) Temperature dependence of magnetization measured with a flux qubit (open circle). 
		The solid line is a fitting to the data. The flux shift is defined as the change from the value at 200 mK.
	} 
	\label{f3}
\end{figure}
The temperature dependence of the magnetization was measured using the magnetometer based on the flux qubit [$I_\mathrm{p}=$ 459 nA and $\Delta/h$ = 1.91 GHz, Fig. \ref{f3}(a)].
Here, the magnetic field was fixed near the operation point of $\Phi = 1/2\Phi_0$, corresponding to $|\mathbf{B}| = $ 0.2 mT.
As shown in Fig. \ref{f3}(b), the magnetization decreased with increasing temperature.
This behavior is consistent with a magnetization that originates from electron spins:
spins are highly polarized if magnetic energy dominates thermal fluctuations.
The experimental results can be fitted by the following function:
\begin{equation}
	f(T) = c_0 + c_1 f_1(T) + c_2 f_2(T), 
\end{equation}
where $c_i (i = 0, 1 , 2)$ are fitting parameters.
$c_0$ is the offset of the magnetometer, while $c_1$ and $c_2$ correspond to the ratio of the impurity amounts.
$f_j (j = 1, 2)$ represents the temperature dependence of the magnetization arising from two different spin species.
Here, $f_1(T)$ is derived from the Hamiltonian of the bismuth impurity [Eq. (\ref{eq3})].
$f_2(T)$ assumes a magnetization from the spin 1/2 system in the substrate with $g\sim2$ [same as Eq. (\ref{eq4})].
The best agreement between experiment and fit was obtained for the parameters $c_1/(c_1+c_2) = 0.873 \pm 0.007$ and $c_2/(c_1+c_2) = 0.127 \pm 0.007$.
These values suggest that impurities in the substrate are dominated by bismuth spins, but a non-negligible number in the spin 1/2 system exist in or on the substrate.
It is worth mentioning that the residual of the fitting increases by at least two times if only bismuth or the spin 1/2 system are considered in the fitting.

\begin{figure}
	\centering
	\includegraphics{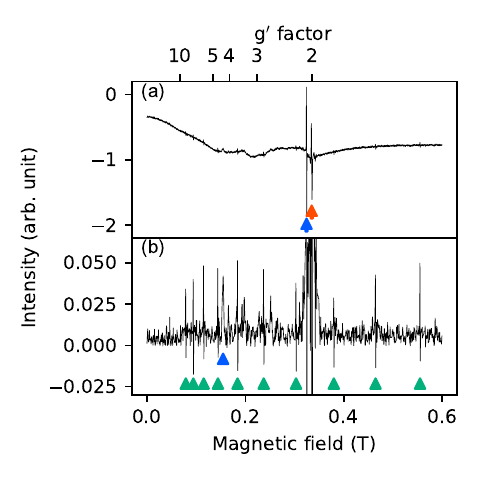}
	\caption{
		Electron spin resonance spectrum obtained by a conventional spectrometer at 20 K.
		(a) Overall spectrum.
		(b) Enlarged view of the spectrum in (a) with background subtraction.
		The background was calculated by using the baseline estimation and denoising using sparsity (BEADS) method \cite{Ning2014}.
		Red, blue, and green arrows indicate peaks for silicon dangling bond, iron, and bismuth, respectively. 
	}
	\label{f4}
\end{figure}

To obtain more insight into the details of the spin 1/2 system, the electron spin resonance (ESR) spectrum of the bismuth-doped silicon substrate without a qubit was measured using a conventional X-band spectrometer, as shown in Fig. \ref{f4}. 
Ten small peaks  with similar intensities appear in the spectrum.
These peaks correspond to transitions between energy levels of bismuth impurities with the same nuclear spin state; i.e., only the electron spin is flipped.

Another prominent peak with $g^\prime \sim 2.006$ is the signature of the silicon dangling bond at the SiO$_2$/Si(100) interface (P$_{\mathrm{b0}}$ center) \cite{Lepine1972, Poindexter1981}.
This is likely one of the origins for the spin 1/2 components found in the magnetometry experiments.
However, since $g^\prime \sim 2$ is the most ordinal component of ESR spectroscopy, other components, e.g. impurity spins located in metal-substrate and metal-air interfaces, may contribute to the magnetization in case of the substrate with a qubit.

The peaks at $g^\prime \sim 4.3$ and 2.07 are attributed to Fe$^{3+}$ ions and Fe$^0$, respectively \cite{Woodbury1960}.
They may have originated from contamination of the silicon wafer during the dicing process.
The contribution of iron impurities to the measured magnetization is negligible because the qubit is sensitive only to the spins near its loop.

The sensitivity of the magnetometry can be evaluated by using a transfer function from magnetization to number of spins.
In this method, noise from the measurement chain is converted into a number of spins, which is defined as a sensitivity \cite{Toida2019}.
From the qubit parameters, the sensitivity of the flux qubit magnetometer is estimated to be 12 spins/$\sqrt{\mathrm{Hz}}$.
This value is about twice as good as a similar previous device \cite{Budoyo2020}.
Compared with that device, the flux qubit used in our experiment had a larger persistent current $I_\mathrm{p}$ by 1.4 times \cite{Budoyo2020}.
The remaining improvement may have originated from changing the readout method from Josephson bifurcation amplifier readout \cite{Budoyo2020} to dispersive readout.

The detection volume of the magnetometry can be easily calculated from the detection area and the effective interaction depth between the spins and the qubit \cite{Marcos2010}.
By multiplying these two parameters, the detection volume is estimated to be 5 $\mu \mathrm{m}^3$ (5 fL).

The sensitivity can be compared to the sensitivity of a SIMS measurement $10^{12}$--$10^{16}$ atoms/cm$^3$ ($=$ 1--10$^4$ atoms/$\mu$m$^3$).
Since $1/f$ flux noise limits the integration time of the flux qubit signal to $\sim$1 second \cite{Toida2019, Budoyo2020}, our current device can detect $\sim$10 spins in a 5 $\mu \mathrm{m}^3$ volume, thus the corresponding volume sensitivity is 2 spins/$\mu$m$^3$ .
This value is comparable to a SIMS measurement.
The sensitivity can be improved by increasing the signal-to-noise ratio of the measurement chain with a quantum limited amplifier, or by optimizing the loop size of the qubit.

Although bismuth impurities were the focus of this paper, the methodology presented herein can naturally be extended to other impurities.
For example, boron, the most abundant accepter in silicon substrate, might limit the qubit coherence time \cite{Zhang2024}.
Since boron impurities in silicon show the quadratic Zeeman effect \cite{Koepf1992}, the magnetic field dependence of the magnetization can provide information to distinguish them from other spin species.

In summary, we successfully detected bismuth impurities in a silicon substrate by using a superconducting flux qubit.
By fitting the temperature dependence of the magnetization, two different spin species were quantitatively identified.
Further improvements in sensitivity are needed to evaluate low density spins.
In the current measurement setup, the signal-to-noise ratio is limited by the first-stage high electron mobility transistor (HEMT) amplifier.
It can be replaced with an amplifier having quantum-limited noise performance, such as a Josephson parametric amplifier or a traveling wave parametric amplifier.
Furthermore, the size of the qubit can be increased to improve the volume sensitivity.
A previous experiment \cite{Toida2019} showed that qubit can be enlarged to at least $\sim$50 $\mu$m$^2$ without its measurements suffering from flux noise; this would lead to it having  $\sim$10 times higher sensitivity.

\begin{acknowledgments}
This work was supported by CREST (Grant No. JPMJCR1774).
The authors thank Taishi Kawabata for fabricating the flux qubit.
\end{acknowledgments}

\section*{DATA AVAILABILITY}
The data that support the findings of this study are available from the corresponding author upon reasonable request.


\begin{thebibliography}{25}%
	\makeatletter
	\providecommand \@ifxundefined [1]{%
		\@ifx{#1\undefined}
	}%
	\providecommand \@ifnum [1]{%
		\ifnum #1\expandafter \@firstoftwo
		\else \expandafter \@secondoftwo
		\fi
	}%
	\providecommand \@ifx [1]{%
		\ifx #1\expandafter \@firstoftwo
		\else \expandafter \@secondoftwo
		\fi
	}%
	\providecommand \natexlab [1]{#1}%
	\providecommand \enquote  [1]{``#1''}%
	\providecommand \bibnamefont  [1]{#1}%
	\providecommand \bibfnamefont [1]{#1}%
	\providecommand \citenamefont [1]{#1}%
	\providecommand \href@noop [0]{\@secondoftwo}%
	\providecommand \href [0]{\begingroup \@sanitize@url \@href}%
	\providecommand \@href[1]{\@@startlink{#1}\@@href}%
	\providecommand \@@href[1]{\endgroup#1\@@endlink}%
	\providecommand \@sanitize@url [0]{\catcode `\\12\catcode `\$12\catcode
		`\&12\catcode `\#12\catcode `\^12\catcode `\_12\catcode `\%12\relax}%
	\providecommand \@@startlink[1]{}%
	\providecommand \@@endlink[0]{}%
	\providecommand \url  [0]{\begingroup\@sanitize@url \@url }%
	\providecommand \@url [1]{\endgroup\@href {#1}{\urlprefix }}%
	\providecommand \urlprefix  [0]{URL }%
	\providecommand \Eprint [0]{\href }%
	\providecommand \doibase [0]{http://dx.doi.org/}%
	\providecommand \selectlanguage [0]{\@gobble}%
	\providecommand \bibinfo  [0]{\@secondoftwo}%
	\providecommand \bibfield  [0]{\@secondoftwo}%
	\providecommand \translation [1]{[#1]}%
	\providecommand \BibitemOpen [0]{}%
	\providecommand \bibitemStop [0]{}%
	\providecommand \bibitemNoStop [0]{.\EOS\space}%
	\providecommand \EOS [0]{\spacefactor3000\relax}%
	\providecommand \BibitemShut  [1]{\csname bibitem#1\endcsname}%
	\let\auto@bib@innerbib\@empty
	\bibitem [{\citenamefont {Arute}\ \emph {et~al.}(2019)\citenamefont {Arute},
		\citenamefont {Arya}, \citenamefont {Babbush}, \citenamefont {Bacon},
		\citenamefont {Bardin}, \citenamefont {Barends}, \citenamefont {Biswas},
		\citenamefont {Boixo}, \citenamefont {Brandao}, \citenamefont {Buell},
		\citenamefont {Burkett}, \citenamefont {Chen}, \citenamefont {Chen},
		\citenamefont {Chiaro}, \citenamefont {Collins}, \citenamefont {Courtney},
		\citenamefont {Dunsworth}, \citenamefont {Farhi}, \citenamefont {Foxen},
		\citenamefont {Fowler}, \citenamefont {Gidney}, \citenamefont {Giustina},
		\citenamefont {Graff}, \citenamefont {Guerin}, \citenamefont {Habegger},
		\citenamefont {Harrigan}, \citenamefont {Hartmann}, \citenamefont {Ho},
		\citenamefont {Hoffmann}, \citenamefont {Huang}, \citenamefont {Humble},
		\citenamefont {Isakov}, \citenamefont {Jeffrey}, \citenamefont {Jiang},
		\citenamefont {Kafri}, \citenamefont {Kechedzhi}, \citenamefont {Kelly},
		\citenamefont {Klimov}, \citenamefont {Knysh}, \citenamefont {Korotkov},
		\citenamefont {Kostritsa}, \citenamefont {Landhuis}, \citenamefont
		{Lindmark}, \citenamefont {Lucero}, \citenamefont {Lyakh}, \citenamefont
		{Mandrà}, \citenamefont {McClean}, \citenamefont {McEwen}, \citenamefont
		{Megrant}, \citenamefont {Mi}, \citenamefont {Michielsen}, \citenamefont
		{Mohseni}, \citenamefont {Mutus}, \citenamefont {Naaman}, \citenamefont
		{Neeley}, \citenamefont {Neill}, \citenamefont {Niu}, \citenamefont {Ostby},
		\citenamefont {Petukhov}, \citenamefont {Platt}, \citenamefont {Quintana},
		\citenamefont {Rieffel}, \citenamefont {Roushan}, \citenamefont {Rubin},
		\citenamefont {Sank}, \citenamefont {Satzinger}, \citenamefont {Smelyanskiy},
		\citenamefont {Sung}, \citenamefont {Trevithick}, \citenamefont
		{Vainsencher}, \citenamefont {Villalonga}, \citenamefont {White},
		\citenamefont {Yao}, \citenamefont {Yeh}, \citenamefont {Zalcman},
		\citenamefont {Neven},\ and\ \citenamefont {Martinis}}]{Arute2019}%
	\BibitemOpen
	\bibfield  {author} {\bibinfo {author} {\bibfnamefont {F.}~\bibnamefont
			{Arute}}, \bibinfo {author} {\bibfnamefont {K.}~\bibnamefont {Arya}},
		\bibinfo {author} {\bibfnamefont {R.}~\bibnamefont {Babbush}}, \bibinfo
		{author} {\bibfnamefont {D.}~\bibnamefont {Bacon}}, \bibinfo {author}
		{\bibfnamefont {J.~C.}\ \bibnamefont {Bardin}}, \bibinfo {author}
		{\bibfnamefont {R.}~\bibnamefont {Barends}}, \bibinfo {author} {\bibfnamefont
			{R.}~\bibnamefont {Biswas}}, \bibinfo {author} {\bibfnamefont
			{S.}~\bibnamefont {Boixo}}, \bibinfo {author} {\bibfnamefont {F.~G. S.~L.}\
			\bibnamefont {Brandao}}, \bibinfo {author} {\bibfnamefont {D.~A.}\
			\bibnamefont {Buell}}, \bibinfo {author} {\bibfnamefont {B.}~\bibnamefont
			{Burkett}}, \bibinfo {author} {\bibfnamefont {Y.}~\bibnamefont {Chen}},
		\bibinfo {author} {\bibfnamefont {Z.}~\bibnamefont {Chen}}, \bibinfo {author}
		{\bibfnamefont {B.}~\bibnamefont {Chiaro}}, \bibinfo {author} {\bibfnamefont
			{R.}~\bibnamefont {Collins}}, \bibinfo {author} {\bibfnamefont
			{W.}~\bibnamefont {Courtney}}, \bibinfo {author} {\bibfnamefont
			{A.}~\bibnamefont {Dunsworth}}, \bibinfo {author} {\bibfnamefont
			{E.}~\bibnamefont {Farhi}}, \bibinfo {author} {\bibfnamefont
			{B.}~\bibnamefont {Foxen}}, \bibinfo {author} {\bibfnamefont
			{A.}~\bibnamefont {Fowler}}, \bibinfo {author} {\bibfnamefont
			{C.}~\bibnamefont {Gidney}}, \bibinfo {author} {\bibfnamefont
			{M.}~\bibnamefont {Giustina}}, \bibinfo {author} {\bibfnamefont
			{R.}~\bibnamefont {Graff}}, \bibinfo {author} {\bibfnamefont
			{K.}~\bibnamefont {Guerin}}, \bibinfo {author} {\bibfnamefont
			{S.}~\bibnamefont {Habegger}}, \bibinfo {author} {\bibfnamefont {M.~P.}\
			\bibnamefont {Harrigan}}, \bibinfo {author} {\bibfnamefont {M.~J.}\
			\bibnamefont {Hartmann}}, \bibinfo {author} {\bibfnamefont {A.}~\bibnamefont
			{Ho}}, \bibinfo {author} {\bibfnamefont {M.}~\bibnamefont {Hoffmann}},
		\bibinfo {author} {\bibfnamefont {T.}~\bibnamefont {Huang}}, \bibinfo
		{author} {\bibfnamefont {T.~S.}\ \bibnamefont {Humble}}, \bibinfo {author}
		{\bibfnamefont {S.~V.}\ \bibnamefont {Isakov}}, \bibinfo {author}
		{\bibfnamefont {E.}~\bibnamefont {Jeffrey}}, \bibinfo {author} {\bibfnamefont
			{Z.}~\bibnamefont {Jiang}}, \bibinfo {author} {\bibfnamefont
			{D.}~\bibnamefont {Kafri}}, \bibinfo {author} {\bibfnamefont
			{K.}~\bibnamefont {Kechedzhi}}, \bibinfo {author} {\bibfnamefont
			{J.}~\bibnamefont {Kelly}}, \bibinfo {author} {\bibfnamefont {P.~V.}\
			\bibnamefont {Klimov}}, \bibinfo {author} {\bibfnamefont {S.}~\bibnamefont
			{Knysh}}, \bibinfo {author} {\bibfnamefont {A.}~\bibnamefont {Korotkov}},
		\bibinfo {author} {\bibfnamefont {F.}~\bibnamefont {Kostritsa}}, \bibinfo
		{author} {\bibfnamefont {D.}~\bibnamefont {Landhuis}}, \bibinfo {author}
		{\bibfnamefont {M.}~\bibnamefont {Lindmark}}, \bibinfo {author}
		{\bibfnamefont {E.}~\bibnamefont {Lucero}}, \bibinfo {author} {\bibfnamefont
			{D.}~\bibnamefont {Lyakh}}, \bibinfo {author} {\bibfnamefont
			{S.}~\bibnamefont {Mandrà}}, \bibinfo {author} {\bibfnamefont {J.~R.}\
			\bibnamefont {McClean}}, \bibinfo {author} {\bibfnamefont {M.}~\bibnamefont
			{McEwen}}, \bibinfo {author} {\bibfnamefont {A.}~\bibnamefont {Megrant}},
		\bibinfo {author} {\bibfnamefont {X.}~\bibnamefont {Mi}}, \bibinfo {author}
		{\bibfnamefont {K.}~\bibnamefont {Michielsen}}, \bibinfo {author}
		{\bibfnamefont {M.}~\bibnamefont {Mohseni}}, \bibinfo {author} {\bibfnamefont
			{J.}~\bibnamefont {Mutus}}, \bibinfo {author} {\bibfnamefont
			{O.}~\bibnamefont {Naaman}}, \bibinfo {author} {\bibfnamefont
			{M.}~\bibnamefont {Neeley}}, \bibinfo {author} {\bibfnamefont
			{C.}~\bibnamefont {Neill}}, \bibinfo {author} {\bibfnamefont {M.~Y.}\
			\bibnamefont {Niu}}, \bibinfo {author} {\bibfnamefont {E.}~\bibnamefont
			{Ostby}}, \bibinfo {author} {\bibfnamefont {A.}~\bibnamefont {Petukhov}},
		\bibinfo {author} {\bibfnamefont {J.~C.}\ \bibnamefont {Platt}}, \bibinfo
		{author} {\bibfnamefont {C.}~\bibnamefont {Quintana}}, \bibinfo {author}
		{\bibfnamefont {E.~G.}\ \bibnamefont {Rieffel}}, \bibinfo {author}
		{\bibfnamefont {P.}~\bibnamefont {Roushan}}, \bibinfo {author} {\bibfnamefont
			{N.~C.}\ \bibnamefont {Rubin}}, \bibinfo {author} {\bibfnamefont
			{D.}~\bibnamefont {Sank}}, \bibinfo {author} {\bibfnamefont {K.~J.}\
			\bibnamefont {Satzinger}}, \bibinfo {author} {\bibfnamefont {V.}~\bibnamefont
			{Smelyanskiy}}, \bibinfo {author} {\bibfnamefont {K.~J.}\ \bibnamefont
			{Sung}}, \bibinfo {author} {\bibfnamefont {M.~D.}\ \bibnamefont
			{Trevithick}}, \bibinfo {author} {\bibfnamefont {A.}~\bibnamefont
			{Vainsencher}}, \bibinfo {author} {\bibfnamefont {B.}~\bibnamefont
			{Villalonga}}, \bibinfo {author} {\bibfnamefont {T.}~\bibnamefont {White}},
		\bibinfo {author} {\bibfnamefont {Z.~J.}\ \bibnamefont {Yao}}, \bibinfo
		{author} {\bibfnamefont {P.}~\bibnamefont {Yeh}}, \bibinfo {author}
		{\bibfnamefont {A.}~\bibnamefont {Zalcman}}, \bibinfo {author} {\bibfnamefont
			{H.}~\bibnamefont {Neven}}, \ and\ \bibinfo {author} {\bibfnamefont {J.~M.}\
			\bibnamefont {Martinis}},\ }\bibfield  {title} {\enquote {\bibinfo {title}
			{Quantum supremacy using a programmable superconducting processor},}\ }\href
	{https://doi.org/10.1038/s41586-019-1666-5} {\bibfield  {journal} {\bibinfo
			{journal} {Nature}\ }\textbf {\bibinfo {volume} {574}},\ \bibinfo {pages}
		{505--510} (\bibinfo {year} {2019})}\BibitemShut {NoStop}%
	\bibitem [{\citenamefont {Kjaergaard}\ \emph {et~al.}(2020)\citenamefont
		{Kjaergaard}, \citenamefont {Schwartz}, \citenamefont {Braumüller},
		\citenamefont {Krantz}, \citenamefont {Wang}, \citenamefont {Gustavsson},\
		and\ \citenamefont {Oliver}}]{Kjaergaard2019}%
	\BibitemOpen
	\bibfield  {author} {\bibinfo {author} {\bibfnamefont {M.}~\bibnamefont
			{Kjaergaard}}, \bibinfo {author} {\bibfnamefont {M.~E.}\ \bibnamefont
			{Schwartz}}, \bibinfo {author} {\bibfnamefont {J.}~\bibnamefont
			{Braumüller}}, \bibinfo {author} {\bibfnamefont {P.}~\bibnamefont {Krantz}},
		\bibinfo {author} {\bibfnamefont {J.~I.-J.}\ \bibnamefont {Wang}}, \bibinfo
		{author} {\bibfnamefont {S.}~\bibnamefont {Gustavsson}}, \ and\ \bibinfo
		{author} {\bibfnamefont {W.~D.}\ \bibnamefont {Oliver}},\ }\bibfield  {title}
	{\enquote {\bibinfo {title} {Superconducting qubits: Current state of
				play},}\ }\href {\doibase 10.1146/annurev-conmatphys-031119-050605}
	{\bibfield  {journal} {\bibinfo  {journal} {Annu. Rev. Condens. Matter
				Phys.}\ }\textbf {\bibinfo {volume} {11}},\ \bibinfo {pages} {369--395}
		(\bibinfo {year} {2020})}\BibitemShut {NoStop}%
	\bibitem [{\citenamefont {Hauke}\ \emph {et~al.}(2020)\citenamefont {Hauke},
		\citenamefont {Katzgraber}, \citenamefont {Lechner}, \citenamefont
		{Nishimori},\ and\ \citenamefont {Oliver}}]{Hauke2020}%
	\BibitemOpen
	\bibfield  {author} {\bibinfo {author} {\bibfnamefont {P.}~\bibnamefont
			{Hauke}}, \bibinfo {author} {\bibfnamefont {H.~G.}\ \bibnamefont
			{Katzgraber}}, \bibinfo {author} {\bibfnamefont {W.}~\bibnamefont {Lechner}},
		\bibinfo {author} {\bibfnamefont {H.}~\bibnamefont {Nishimori}}, \ and\
		\bibinfo {author} {\bibfnamefont {W.~D.}\ \bibnamefont {Oliver}},\ }\bibfield
	{title} {\enquote {\bibinfo {title} {Perspectives of quantum annealing:
				methods and implementations},}\ }\href {\doibase 10.1088/1361-6633/ab85b8}
	{\bibfield  {journal} {\bibinfo  {journal} {Rep. Prog. Phys.}\ }\textbf
		{\bibinfo {volume} {83}},\ \bibinfo {pages} {054401} (\bibinfo {year}
		{2020})}\BibitemShut {NoStop}%
	\bibitem [{\citenamefont {King}\ \emph {et~al.}()\citenamefont {King},
		\citenamefont {Nocera}, \citenamefont {Rams}, \citenamefont {Dziarmaga},
		\citenamefont {Wiersema}, \citenamefont {Bernoudy}, \citenamefont {Raymond},
		\citenamefont {Kaushal}, \citenamefont {Heinsdorf}, \citenamefont {Harris},
		\citenamefont {Boothby}, \citenamefont {Altomare}, \citenamefont {Berkley},
		\citenamefont {Boschnak}, \citenamefont {Chern}, \citenamefont {Christiani},
		\citenamefont {Cibere}, \citenamefont {Connor}, \citenamefont {Dehn},
		\citenamefont {Deshpande}, \citenamefont {Ejtemaee}, \citenamefont {Farré},
		\citenamefont {Hamer}, \citenamefont {Hoskinson}, \citenamefont {Huang},
		\citenamefont {Johnson}, \citenamefont {Kortas}, \citenamefont {Ladizinsky},
		\citenamefont {Lai}, \citenamefont {Lanting}, \citenamefont {Li},
		\citenamefont {MacDonald}, \citenamefont {Marsden}, \citenamefont {McGeoch},
		\citenamefont {Molavi}, \citenamefont {Neufeld}, \citenamefont {Norouzpour},
		\citenamefont {Oh}, \citenamefont {Pasvolsky}, \citenamefont {Poitras},
		\citenamefont {Poulin-Lamarre}, \citenamefont {Prescott}, \citenamefont
		{Reis}, \citenamefont {Rich}, \citenamefont {Samani}, \citenamefont
		{Sheldan}, \citenamefont {Smirnov}, \citenamefont {Sterpka}, \citenamefont
		{Clavera}, \citenamefont {Tsai}, \citenamefont {Volkmann}, \citenamefont
		{Whiticar}, \citenamefont {Whittaker}, \citenamefont {Wilkinson},
		\citenamefont {Yao}, \citenamefont {Yi}, \citenamefont {Sandvik},
		\citenamefont {Alvarez}, \citenamefont {Melko}, \citenamefont {Carrasquilla},
		\citenamefont {Franz},\ and\ \citenamefont {Amin}}]{King2024}%
	\BibitemOpen
	\bibfield  {author} {\bibinfo {author} {\bibfnamefont {A.~D.}\ \bibnamefont
			{King}}, \bibinfo {author} {\bibfnamefont {A.}~\bibnamefont {Nocera}},
		\bibinfo {author} {\bibfnamefont {M.~M.}\ \bibnamefont {Rams}}, \bibinfo
		{author} {\bibfnamefont {J.}~\bibnamefont {Dziarmaga}}, \bibinfo {author}
		{\bibfnamefont {R.}~\bibnamefont {Wiersema}}, \bibinfo {author}
		{\bibfnamefont {W.}~\bibnamefont {Bernoudy}}, \bibinfo {author}
		{\bibfnamefont {J.}~\bibnamefont {Raymond}}, \bibinfo {author} {\bibfnamefont
			{N.}~\bibnamefont {Kaushal}}, \bibinfo {author} {\bibfnamefont
			{N.}~\bibnamefont {Heinsdorf}}, \bibinfo {author} {\bibfnamefont
			{R.}~\bibnamefont {Harris}}, \bibinfo {author} {\bibfnamefont
			{K.}~\bibnamefont {Boothby}}, \bibinfo {author} {\bibfnamefont
			{F.}~\bibnamefont {Altomare}}, \bibinfo {author} {\bibfnamefont {A.~J.}\
			\bibnamefont {Berkley}}, \bibinfo {author} {\bibfnamefont {M.}~\bibnamefont
			{Boschnak}}, \bibinfo {author} {\bibfnamefont {K.}~\bibnamefont {Chern}},
		\bibinfo {author} {\bibfnamefont {H.}~\bibnamefont {Christiani}}, \bibinfo
		{author} {\bibfnamefont {S.}~\bibnamefont {Cibere}}, \bibinfo {author}
		{\bibfnamefont {J.}~\bibnamefont {Connor}}, \bibinfo {author} {\bibfnamefont
			{M.~H.}\ \bibnamefont {Dehn}}, \bibinfo {author} {\bibfnamefont
			{R.}~\bibnamefont {Deshpande}}, \bibinfo {author} {\bibfnamefont
			{S.}~\bibnamefont {Ejtemaee}}, \bibinfo {author} {\bibfnamefont
			{P.}~\bibnamefont {Farré}}, \bibinfo {author} {\bibfnamefont
			{K.}~\bibnamefont {Hamer}}, \bibinfo {author} {\bibfnamefont
			{E.}~\bibnamefont {Hoskinson}}, \bibinfo {author} {\bibfnamefont
			{S.}~\bibnamefont {Huang}}, \bibinfo {author} {\bibfnamefont {M.~W.}\
			\bibnamefont {Johnson}}, \bibinfo {author} {\bibfnamefont {S.}~\bibnamefont
			{Kortas}}, \bibinfo {author} {\bibfnamefont {E.}~\bibnamefont {Ladizinsky}},
		\bibinfo {author} {\bibfnamefont {T.}~\bibnamefont {Lai}}, \bibinfo {author}
		{\bibfnamefont {T.}~\bibnamefont {Lanting}}, \bibinfo {author} {\bibfnamefont
			{R.}~\bibnamefont {Li}}, \bibinfo {author} {\bibfnamefont {A.~J.~R.}\
			\bibnamefont {MacDonald}}, \bibinfo {author} {\bibfnamefont {G.}~\bibnamefont
			{Marsden}}, \bibinfo {author} {\bibfnamefont {C.~C.}\ \bibnamefont
			{McGeoch}}, \bibinfo {author} {\bibfnamefont {R.}~\bibnamefont {Molavi}},
		\bibinfo {author} {\bibfnamefont {R.}~\bibnamefont {Neufeld}}, \bibinfo
		{author} {\bibfnamefont {M.}~\bibnamefont {Norouzpour}}, \bibinfo {author}
		{\bibfnamefont {T.}~\bibnamefont {Oh}}, \bibinfo {author} {\bibfnamefont
			{J.}~\bibnamefont {Pasvolsky}}, \bibinfo {author} {\bibfnamefont
			{P.}~\bibnamefont {Poitras}}, \bibinfo {author} {\bibfnamefont
			{G.}~\bibnamefont {Poulin-Lamarre}}, \bibinfo {author} {\bibfnamefont
			{T.}~\bibnamefont {Prescott}}, \bibinfo {author} {\bibfnamefont
			{M.}~\bibnamefont {Reis}}, \bibinfo {author} {\bibfnamefont {C.}~\bibnamefont
			{Rich}}, \bibinfo {author} {\bibfnamefont {M.}~\bibnamefont {Samani}},
		\bibinfo {author} {\bibfnamefont {B.}~\bibnamefont {Sheldan}}, \bibinfo
		{author} {\bibfnamefont {A.}~\bibnamefont {Smirnov}}, \bibinfo {author}
		{\bibfnamefont {E.}~\bibnamefont {Sterpka}}, \bibinfo {author} {\bibfnamefont
			{B.~T.}\ \bibnamefont {Clavera}}, \bibinfo {author} {\bibfnamefont
			{N.}~\bibnamefont {Tsai}}, \bibinfo {author} {\bibfnamefont {M.}~\bibnamefont
			{Volkmann}}, \bibinfo {author} {\bibfnamefont {A.}~\bibnamefont {Whiticar}},
		\bibinfo {author} {\bibfnamefont {J.~D.}\ \bibnamefont {Whittaker}}, \bibinfo
		{author} {\bibfnamefont {W.}~\bibnamefont {Wilkinson}}, \bibinfo {author}
		{\bibfnamefont {J.}~\bibnamefont {Yao}}, \bibinfo {author} {\bibfnamefont
			{T.~J.}\ \bibnamefont {Yi}}, \bibinfo {author} {\bibfnamefont {A.~W.}\
			\bibnamefont {Sandvik}}, \bibinfo {author} {\bibfnamefont {G.}~\bibnamefont
			{Alvarez}}, \bibinfo {author} {\bibfnamefont {R.~G.}\ \bibnamefont {Melko}},
		\bibinfo {author} {\bibfnamefont {J.}~\bibnamefont {Carrasquilla}}, \bibinfo
		{author} {\bibfnamefont {M.}~\bibnamefont {Franz}}, \ and\ \bibinfo {author}
		{\bibfnamefont {M.~H.}\ \bibnamefont {Amin}},\ }\bibfield  {title} {\enquote
		{\bibinfo {title} {Computational supremacy in quantum simulation},}\
	}\href@noop {} {\ }\Eprint {http://arxiv.org/abs/2403.00910}
	{arXiv:2403.00910 [quant-ph]} \BibitemShut {NoStop}%
	\bibitem [{\citenamefont {Woods}\ \emph {et~al.}(2019)\citenamefont {Woods},
		\citenamefont {Calusine}, \citenamefont {Melville}, \citenamefont {Sevi},
		\citenamefont {Golden}, \citenamefont {Kim}, \citenamefont {Rosenberg},
		\citenamefont {Yoder},\ and\ \citenamefont {Oliver}}]{Woods2019}%
	\BibitemOpen
	\bibfield  {author} {\bibinfo {author} {\bibfnamefont {W.}~\bibnamefont
			{Woods}}, \bibinfo {author} {\bibfnamefont {G.}~\bibnamefont {Calusine}},
		\bibinfo {author} {\bibfnamefont {A.}~\bibnamefont {Melville}}, \bibinfo
		{author} {\bibfnamefont {A.}~\bibnamefont {Sevi}}, \bibinfo {author}
		{\bibfnamefont {E.}~\bibnamefont {Golden}}, \bibinfo {author} {\bibfnamefont
			{D.}~\bibnamefont {Kim}}, \bibinfo {author} {\bibfnamefont {D.}~\bibnamefont
			{Rosenberg}}, \bibinfo {author} {\bibfnamefont {J.}~\bibnamefont {Yoder}}, \
		and\ \bibinfo {author} {\bibfnamefont {W.}~\bibnamefont {Oliver}},\
	}\bibfield  {title} {\enquote {\bibinfo {title} {Determining interface
				dielectric losses in superconducting coplanar-waveguide resonators},}\ }\href
	{\doibase 10.1103/PhysRevApplied.12.014012} {\bibfield  {journal} {\bibinfo
			{journal} {Phys. Rev. Applied}\ }\textbf {\bibinfo {volume} {12}},\ \bibinfo
		{pages} {014012} (\bibinfo {year} {2019})}\BibitemShut {NoStop}%
	\bibitem [{\citenamefont {Siddiqi}(2021)}]{Siddiqi2021}%
	\BibitemOpen
	\bibfield  {author} {\bibinfo {author} {\bibfnamefont {I.}~\bibnamefont
			{Siddiqi}},\ }\bibfield  {title} {\enquote {\bibinfo {title} {Engineering
				high-coherence superconducting qubits},}\ }\href {\doibase
		10.1038/s41578-021-00370-4} {\bibfield  {journal} {\bibinfo  {journal} {Nat.
				Rev. Mater.}\ }\textbf {\bibinfo {volume} {6}},\ \bibinfo {pages} {875--891}
		(\bibinfo {year} {2021})}\BibitemShut {NoStop}%
	\bibitem [{\citenamefont {Wang}\ \emph {et~al.}(2022)\citenamefont {Wang},
		\citenamefont {Li}, \citenamefont {Xu}, \citenamefont {Li}, \citenamefont
		{Wang}, \citenamefont {Yang}, \citenamefont {Mi}, \citenamefont {Liang},
		\citenamefont {Su}, \citenamefont {Yang}, \citenamefont {Wang}, \citenamefont
		{Wang}, \citenamefont {Li}, \citenamefont {Chen}, \citenamefont {Li},
		\citenamefont {Linghu}, \citenamefont {Han}, \citenamefont {Zhang},
		\citenamefont {Feng}, \citenamefont {Song}, \citenamefont {Ma}, \citenamefont
		{Zhang}, \citenamefont {Wang}, \citenamefont {Zhao}, \citenamefont {Liu},
		\citenamefont {Xue}, \citenamefont {Jin},\ and\ \citenamefont
		{Yu}}]{Wang2022}%
	\BibitemOpen
	\bibfield  {author} {\bibinfo {author} {\bibfnamefont {C.}~\bibnamefont
			{Wang}}, \bibinfo {author} {\bibfnamefont {X.}~\bibnamefont {Li}}, \bibinfo
		{author} {\bibfnamefont {H.}~\bibnamefont {Xu}}, \bibinfo {author}
		{\bibfnamefont {Z.}~\bibnamefont {Li}}, \bibinfo {author} {\bibfnamefont
			{J.}~\bibnamefont {Wang}}, \bibinfo {author} {\bibfnamefont {Z.}~\bibnamefont
			{Yang}}, \bibinfo {author} {\bibfnamefont {Z.}~\bibnamefont {Mi}}, \bibinfo
		{author} {\bibfnamefont {X.}~\bibnamefont {Liang}}, \bibinfo {author}
		{\bibfnamefont {T.}~\bibnamefont {Su}}, \bibinfo {author} {\bibfnamefont
			{C.}~\bibnamefont {Yang}}, \bibinfo {author} {\bibfnamefont {G.}~\bibnamefont
			{Wang}}, \bibinfo {author} {\bibfnamefont {W.}~\bibnamefont {Wang}}, \bibinfo
		{author} {\bibfnamefont {Y.}~\bibnamefont {Li}}, \bibinfo {author}
		{\bibfnamefont {M.}~\bibnamefont {Chen}}, \bibinfo {author} {\bibfnamefont
			{C.}~\bibnamefont {Li}}, \bibinfo {author} {\bibfnamefont {K.}~\bibnamefont
			{Linghu}}, \bibinfo {author} {\bibfnamefont {J.}~\bibnamefont {Han}},
		\bibinfo {author} {\bibfnamefont {Y.}~\bibnamefont {Zhang}}, \bibinfo
		{author} {\bibfnamefont {Y.}~\bibnamefont {Feng}}, \bibinfo {author}
		{\bibfnamefont {Y.}~\bibnamefont {Song}}, \bibinfo {author} {\bibfnamefont
			{T.}~\bibnamefont {Ma}}, \bibinfo {author} {\bibfnamefont {J.}~\bibnamefont
			{Zhang}}, \bibinfo {author} {\bibfnamefont {R.}~\bibnamefont {Wang}},
		\bibinfo {author} {\bibfnamefont {P.}~\bibnamefont {Zhao}}, \bibinfo {author}
		{\bibfnamefont {W.}~\bibnamefont {Liu}}, \bibinfo {author} {\bibfnamefont
			{G.}~\bibnamefont {Xue}}, \bibinfo {author} {\bibfnamefont {Y.}~\bibnamefont
			{Jin}}, \ and\ \bibinfo {author} {\bibfnamefont {H.}~\bibnamefont {Yu}},\
	}\bibfield  {title} {\enquote {\bibinfo {title} {Towards practical quantum
				computers: transmon qubit with a lifetime approaching 0.5 milliseconds},}\
	}\href {\doibase 10.1038/s41534-021-00510-2} {\bibfield  {journal} {\bibinfo
			{journal} {npj Quantum Inf.}\ }\textbf {\bibinfo {volume} {8}},\ \bibinfo
		{pages} {3} (\bibinfo {year} {2022})}\BibitemShut {NoStop}%
	\bibitem [{\citenamefont {Zhang}\ \emph {et~al.}()\citenamefont {Zhang},
		\citenamefont {Godeneli}, \citenamefont {He}, \citenamefont {Odeh},
		\citenamefont {Zhou}, \citenamefont {Meesala},\ and\ \citenamefont
		{Sipahigil}}]{Zhang2024}%
	\BibitemOpen
	\bibfield  {author} {\bibinfo {author} {\bibfnamefont {Z.-H.}\ \bibnamefont
			{Zhang}}, \bibinfo {author} {\bibfnamefont {K.}~\bibnamefont {Godeneli}},
		\bibinfo {author} {\bibfnamefont {J.}~\bibnamefont {He}}, \bibinfo {author}
		{\bibfnamefont {M.}~\bibnamefont {Odeh}}, \bibinfo {author} {\bibfnamefont
			{H.}~\bibnamefont {Zhou}}, \bibinfo {author} {\bibfnamefont {S.}~\bibnamefont
			{Meesala}}, \ and\ \bibinfo {author} {\bibfnamefont {A.}~\bibnamefont
			{Sipahigil}},\ }\bibfield  {title} {\enquote {\bibinfo {title}
			{Acceptor-induced bulk dielectric loss in superconducting circuits on
				silicon},}\ }\href@noop {} {\ }\Eprint {http://arxiv.org/abs/2402.17155}
	{arXiv:2402.17155 [quant-ph]} \BibitemShut {NoStop}%
	\bibitem [{\citenamefont {Bienfait}\ \emph {et~al.}(2016)\citenamefont
		{Bienfait}, \citenamefont {Pla}, \citenamefont {Kubo}, \citenamefont {Stern},
		\citenamefont {Zhou}, \citenamefont {Lo}, \citenamefont {C.~D.},
		\citenamefont {Schenkel}, \citenamefont {Thewalt}, \citenamefont {Vion},
		\citenamefont {Esteve}, \citenamefont {Julsgaard}, \citenamefont {M{\o}lmer},
		\citenamefont {Morton},\ and\ \citenamefont {Bertet}}]{Bienfait2016}%
	\BibitemOpen
	\bibfield  {author} {\bibinfo {author} {\bibfnamefont {A.}~\bibnamefont
			{Bienfait}}, \bibinfo {author} {\bibfnamefont {J.~J.}\ \bibnamefont {Pla}},
		\bibinfo {author} {\bibfnamefont {Y.}~\bibnamefont {Kubo}}, \bibinfo {author}
		{\bibfnamefont {M.}~\bibnamefont {Stern}}, \bibinfo {author} {\bibfnamefont
			{X.}~\bibnamefont {Zhou}}, \bibinfo {author} {\bibfnamefont {C.~C.}\
			\bibnamefont {Lo}}, \bibinfo {author} {\bibfnamefont {W.}~\bibnamefont
			{C.~D.}}, \bibinfo {author} {\bibfnamefont {T.}~\bibnamefont {Schenkel}},
		\bibinfo {author} {\bibfnamefont {M.~L.~W.}\ \bibnamefont {Thewalt}},
		\bibinfo {author} {\bibfnamefont {D.}~\bibnamefont {Vion}}, \bibinfo {author}
		{\bibfnamefont {D.}~\bibnamefont {Esteve}}, \bibinfo {author} {\bibfnamefont
			{B.}~\bibnamefont {Julsgaard}}, \bibinfo {author} {\bibfnamefont
			{K.}~\bibnamefont {M{\o}lmer}}, \bibinfo {author} {\bibfnamefont {J.~J.~L.}\
			\bibnamefont {Morton}}, \ and\ \bibinfo {author} {\bibfnamefont
			{P.}~\bibnamefont {Bertet}},\ }\bibfield  {title} {\enquote {\bibinfo {title}
			{Reaching the quantum limit of sensitivity in electron spin resonance},}\
	}\href {http://dx.doi.org/10.1038/nnano.2015.282} {\bibfield  {journal}
		{\bibinfo  {journal} {Nat. Nanotechnol.}\ }\textbf {\bibinfo {volume} {11}},\
		\bibinfo {pages} {253--257} (\bibinfo {year} {2016})}\BibitemShut {NoStop}%
	\bibitem [{\citenamefont {Probst}\ \emph {et~al.}(2017)\citenamefont {Probst},
		\citenamefont {Bienfait}, \citenamefont {Campagne-Ibarcq}, \citenamefont
		{Pla}, \citenamefont {Albanese}, \citenamefont {Barbosa}, \citenamefont
		{Schenkel}, \citenamefont {Vion}, \citenamefont {Esteve}, \citenamefont
		{M{\o}lmer}, \citenamefont {Morton}, \citenamefont {Heeres},\ and\
		\citenamefont {Bertet}}]{Probst2017a}%
	\BibitemOpen
	\bibfield  {author} {\bibinfo {author} {\bibfnamefont {S.}~\bibnamefont
			{Probst}}, \bibinfo {author} {\bibfnamefont {A.}~\bibnamefont {Bienfait}},
		\bibinfo {author} {\bibfnamefont {P.}~\bibnamefont {Campagne-Ibarcq}},
		\bibinfo {author} {\bibfnamefont {J.~J.}\ \bibnamefont {Pla}}, \bibinfo
		{author} {\bibfnamefont {B.}~\bibnamefont {Albanese}}, \bibinfo {author}
		{\bibfnamefont {J.~F. D.~S.}\ \bibnamefont {Barbosa}}, \bibinfo {author}
		{\bibfnamefont {T.}~\bibnamefont {Schenkel}}, \bibinfo {author}
		{\bibfnamefont {D.}~\bibnamefont {Vion}}, \bibinfo {author} {\bibfnamefont
			{D.}~\bibnamefont {Esteve}}, \bibinfo {author} {\bibfnamefont
			{K.}~\bibnamefont {M{\o}lmer}}, \bibinfo {author} {\bibfnamefont {J.~J.~L.}\
			\bibnamefont {Morton}}, \bibinfo {author} {\bibfnamefont {R.}~\bibnamefont
			{Heeres}}, \ and\ \bibinfo {author} {\bibfnamefont {P.}~\bibnamefont
			{Bertet}},\ }\bibfield  {title} {\enquote {\bibinfo {title}
			{Inductive-detection electron-spin resonance spectroscopy with 65
				spins/$\sqrt\mathrm{{Hz}}$ sensitivity},}\ }\href {\doibase
		10.1063/1.5002540} {\bibfield  {journal} {\bibinfo  {journal} {Appl. Phys.
				Lett.}\ }\textbf {\bibinfo {volume} {111}},\ \bibinfo {pages} {202604}
		(\bibinfo {year} {2017})}\BibitemShut {NoStop}%
	\bibitem [{\citenamefont {Albertinale}\ \emph {et~al.}(2021)\citenamefont
		{Albertinale}, \citenamefont {Balembois}, \citenamefont {Billaud},
		\citenamefont {Ranjan}, \citenamefont {Flanigan}, \citenamefont {Schenkel},
		\citenamefont {Estève}, \citenamefont {Vion}, \citenamefont {Bertet},\ and\
		\citenamefont {Flurin}}]{Albertinale2021}%
	\BibitemOpen
	\bibfield  {author} {\bibinfo {author} {\bibfnamefont {E.}~\bibnamefont
			{Albertinale}}, \bibinfo {author} {\bibfnamefont {L.}~\bibnamefont
			{Balembois}}, \bibinfo {author} {\bibfnamefont {E.}~\bibnamefont {Billaud}},
		\bibinfo {author} {\bibfnamefont {V.}~\bibnamefont {Ranjan}}, \bibinfo
		{author} {\bibfnamefont {D.}~\bibnamefont {Flanigan}}, \bibinfo {author}
		{\bibfnamefont {T.}~\bibnamefont {Schenkel}}, \bibinfo {author}
		{\bibfnamefont {D.}~\bibnamefont {Estève}}, \bibinfo {author} {\bibfnamefont
			{D.}~\bibnamefont {Vion}}, \bibinfo {author} {\bibfnamefont {P.}~\bibnamefont
			{Bertet}}, \ and\ \bibinfo {author} {\bibfnamefont {E.}~\bibnamefont
			{Flurin}},\ }\bibfield  {title} {\enquote {\bibinfo {title} {Detecting spins
				by their fluorescence with a microwave photon counter},}\ }\href {\doibase
		10.1038/s41586-021-04076-z} {\bibfield  {journal} {\bibinfo  {journal}
			{Nature}\ }\textbf {\bibinfo {volume} {600}},\ \bibinfo {pages} {434--438}
		(\bibinfo {year} {2021})}\BibitemShut {NoStop}%
	\bibitem [{\citenamefont {Wang}\ \emph {et~al.}(2023)\citenamefont {Wang},
		\citenamefont {Balembois}, \citenamefont {Rančić}, \citenamefont {Billaud},
		\citenamefont {Le~Dantec}, \citenamefont {Ferrier}, \citenamefont {Goldner},
		\citenamefont {Bertaina}, \citenamefont {Chanelière}, \citenamefont
		{Esteve}, \citenamefont {Vion}, \citenamefont {Bertet},\ and\ \citenamefont
		{Flurin}}]{Wang2023}%
	\BibitemOpen
	\bibfield  {author} {\bibinfo {author} {\bibfnamefont {Z.}~\bibnamefont
			{Wang}}, \bibinfo {author} {\bibfnamefont {L.}~\bibnamefont {Balembois}},
		\bibinfo {author} {\bibfnamefont {M.}~\bibnamefont {Rančić}}, \bibinfo
		{author} {\bibfnamefont {E.}~\bibnamefont {Billaud}}, \bibinfo {author}
		{\bibfnamefont {M.}~\bibnamefont {Le~Dantec}}, \bibinfo {author}
		{\bibfnamefont {A.}~\bibnamefont {Ferrier}}, \bibinfo {author} {\bibfnamefont
			{P.}~\bibnamefont {Goldner}}, \bibinfo {author} {\bibfnamefont
			{S.}~\bibnamefont {Bertaina}}, \bibinfo {author} {\bibfnamefont
			{T.}~\bibnamefont {Chanelière}}, \bibinfo {author} {\bibfnamefont
			{D.}~\bibnamefont {Esteve}}, \bibinfo {author} {\bibfnamefont
			{D.}~\bibnamefont {Vion}}, \bibinfo {author} {\bibfnamefont {P.}~\bibnamefont
			{Bertet}}, \ and\ \bibinfo {author} {\bibfnamefont {E.}~\bibnamefont
			{Flurin}},\ }\bibfield  {title} {\enquote {\bibinfo {title} {Single-electron
				spin resonance detection by microwave photon counting},}\ }\href {\doibase
		10.1038/s41586-023-06097-2} {\bibfield  {journal} {\bibinfo  {journal}
			{Nature}\ }\textbf {\bibinfo {volume} {619}},\ \bibinfo {pages} {276--281}
		(\bibinfo {year} {2023})}\BibitemShut {NoStop}%
	\bibitem [{\citenamefont {Toida}\ \emph {et~al.}(2019)\citenamefont {Toida},
		\citenamefont {Matsuzaki}, \citenamefont {Kakuyanagi}, \citenamefont {Zhu},
		\citenamefont {Munro}, \citenamefont {Yamaguchi},\ and\ \citenamefont
		{Saito}}]{Toida2019}%
	\BibitemOpen
	\bibfield  {author} {\bibinfo {author} {\bibfnamefont {H.}~\bibnamefont
			{Toida}}, \bibinfo {author} {\bibfnamefont {Y.}~\bibnamefont {Matsuzaki}},
		\bibinfo {author} {\bibfnamefont {K.}~\bibnamefont {Kakuyanagi}}, \bibinfo
		{author} {\bibfnamefont {X.}~\bibnamefont {Zhu}}, \bibinfo {author}
		{\bibfnamefont {W.~J.}\ \bibnamefont {Munro}}, \bibinfo {author}
		{\bibfnamefont {H.}~\bibnamefont {Yamaguchi}}, \ and\ \bibinfo {author}
		{\bibfnamefont {S.}~\bibnamefont {Saito}},\ }\bibfield  {title} {\enquote
		{\bibinfo {title} {Electron paramagnetic resonance spectroscopy using a
				single artificial atom},}\ }\href {https://doi.org/10.1038/s42005-019-0133-9}
	{\bibfield  {journal} {\bibinfo  {journal} {Commun. Phys.}\ }\textbf
		{\bibinfo {volume} {2}},\ \bibinfo {pages} {33} (\bibinfo {year}
		{2019})}\BibitemShut {NoStop}%
	\bibitem [{\citenamefont {Budoyo}\ \emph {et~al.}(2020)\citenamefont {Budoyo},
		\citenamefont {Kakuyanagi}, \citenamefont {Toida}, \citenamefont
		{Matsuzaki},\ and\ \citenamefont {Saito}}]{Budoyo2020}%
	\BibitemOpen
	\bibfield  {author} {\bibinfo {author} {\bibfnamefont {R.~P.}\ \bibnamefont
			{Budoyo}}, \bibinfo {author} {\bibfnamefont {K.}~\bibnamefont {Kakuyanagi}},
		\bibinfo {author} {\bibfnamefont {H.}~\bibnamefont {Toida}}, \bibinfo
		{author} {\bibfnamefont {Y.}~\bibnamefont {Matsuzaki}}, \ and\ \bibinfo
		{author} {\bibfnamefont {S.}~\bibnamefont {Saito}},\ }\bibfield  {title}
	{\enquote {\bibinfo {title} {Electron spin resonance with up to 20 spin
				sensitivity measured using a superconducting flux qubit},}\ }\href {\doibase
		10.1063/1.5144722} {\bibfield  {journal} {\bibinfo  {journal} {Appl. Phys.
				Lett.}\ }\textbf {\bibinfo {volume} {116}},\ \bibinfo {pages} {194001}
		(\bibinfo {year} {2020})}\BibitemShut {NoStop}%
	\bibitem [{\citenamefont {Toida}\ \emph {et~al.}(2023)\citenamefont {Toida},
		\citenamefont {Sakai}, \citenamefont {Teshima}, \citenamefont {Hori},
		\citenamefont {Kakuyanagi}, \citenamefont {Mahboob}, \citenamefont {Ono},\
		and\ \citenamefont {Saito}}]{Toida2023}%
	\BibitemOpen
	\bibfield  {author} {\bibinfo {author} {\bibfnamefont {H.}~\bibnamefont
			{Toida}}, \bibinfo {author} {\bibfnamefont {K.}~\bibnamefont {Sakai}},
		\bibinfo {author} {\bibfnamefont {T.~F.}\ \bibnamefont {Teshima}}, \bibinfo
		{author} {\bibfnamefont {M.}~\bibnamefont {Hori}}, \bibinfo {author}
		{\bibfnamefont {K.}~\bibnamefont {Kakuyanagi}}, \bibinfo {author}
		{\bibfnamefont {I.}~\bibnamefont {Mahboob}}, \bibinfo {author} {\bibfnamefont
			{Y.}~\bibnamefont {Ono}}, \ and\ \bibinfo {author} {\bibfnamefont
			{S.}~\bibnamefont {Saito}},\ }\bibfield  {title} {\enquote {\bibinfo {title}
			{Magnetometry of neurons using a superconducting qubit},}\ }\href {\doibase
		10.1038/s42005-023-01133-z} {\bibfield  {journal} {\bibinfo  {journal}
			{Commun. Phys.}\ }\textbf {\bibinfo {volume} {6}},\ \bibinfo {pages} {19}
		(\bibinfo {year} {2023})}\BibitemShut {NoStop}%
	\bibitem [{\citenamefont {You}\ \emph {et~al.}(2007)\citenamefont {You},
		\citenamefont {Hu}, \citenamefont {Ashhab},\ and\ \citenamefont
		{Nori}}]{You2007}%
	\BibitemOpen
	\bibfield  {author} {\bibinfo {author} {\bibfnamefont {J.~Q.}\ \bibnamefont
			{You}}, \bibinfo {author} {\bibfnamefont {X.}~\bibnamefont {Hu}}, \bibinfo
		{author} {\bibfnamefont {S.}~\bibnamefont {Ashhab}}, \ and\ \bibinfo {author}
		{\bibfnamefont {F.}~\bibnamefont {Nori}},\ }\bibfield  {title} {\enquote
		{\bibinfo {title} {Low-decoherence flux qubit},}\ }\href {\doibase
		10.1103/PhysRevB.75.140515} {\bibfield  {journal} {\bibinfo  {journal} {Phys.
				Rev. B}\ }\textbf {\bibinfo {volume} {75}},\ \bibinfo {pages} {140515}
		(\bibinfo {year} {2007})}\BibitemShut {NoStop}%
	\bibitem [{\citenamefont {Yan}\ \emph {et~al.}(2016)\citenamefont {Yan},
		\citenamefont {Gustavsson}, \citenamefont {Kamal}, \citenamefont {Birenbaum},
		\citenamefont {Sears}, \citenamefont {Hover}, \citenamefont {Gudmundsen},
		\citenamefont {Rosenberg}, \citenamefont {Samach}, \citenamefont {Weber},
		\citenamefont {Yoder}, \citenamefont {Orlando}, \citenamefont {Clarke},
		\citenamefont {Kerman},\ and\ \citenamefont {Oliver}}]{Yan2016}%
	\BibitemOpen
	\bibfield  {author} {\bibinfo {author} {\bibfnamefont {F.}~\bibnamefont
			{Yan}}, \bibinfo {author} {\bibfnamefont {S.}~\bibnamefont {Gustavsson}},
		\bibinfo {author} {\bibfnamefont {A.}~\bibnamefont {Kamal}}, \bibinfo
		{author} {\bibfnamefont {J.}~\bibnamefont {Birenbaum}}, \bibinfo {author}
		{\bibfnamefont {A.~P.}\ \bibnamefont {Sears}}, \bibinfo {author}
		{\bibfnamefont {D.}~\bibnamefont {Hover}}, \bibinfo {author} {\bibfnamefont
			{T.~J.}\ \bibnamefont {Gudmundsen}}, \bibinfo {author} {\bibfnamefont
			{D.}~\bibnamefont {Rosenberg}}, \bibinfo {author} {\bibfnamefont
			{G.}~\bibnamefont {Samach}}, \bibinfo {author} {\bibfnamefont
			{S.}~\bibnamefont {Weber}}, \bibinfo {author} {\bibfnamefont {J.~L.}\
			\bibnamefont {Yoder}}, \bibinfo {author} {\bibfnamefont {T.~P.}\ \bibnamefont
			{Orlando}}, \bibinfo {author} {\bibfnamefont {J.}~\bibnamefont {Clarke}},
		\bibinfo {author} {\bibfnamefont {A.~J.}\ \bibnamefont {Kerman}}, \ and\
		\bibinfo {author} {\bibfnamefont {W.~D.}\ \bibnamefont {Oliver}},\ }\bibfield
	{title} {\enquote {\bibinfo {title} {The flux qubit revisited to enhance
				coherence and reproducibility},}\ }\href
	{http://dx.doi.org/10.1038/ncomms12964} {\bibfield  {journal} {\bibinfo
			{journal} {Nat. Commun.}\ }\textbf {\bibinfo {volume} {7}},\ \bibinfo {pages}
		{12964} (\bibinfo {year} {2016})}\BibitemShut {NoStop}%
	\bibitem [{\citenamefont {Weis}\ \emph {et~al.}(2012)\citenamefont {Weis},
		\citenamefont {Lo}, \citenamefont {Lang}, \citenamefont {Tyryshkin},
		\citenamefont {George}, \citenamefont {Yu}, \citenamefont {Bokor},
		\citenamefont {Lyon}, \citenamefont {Morton},\ and\ \citenamefont
		{Schenkel}}]{Weis2012}%
	\BibitemOpen
	\bibfield  {author} {\bibinfo {author} {\bibfnamefont {C.~D.}\ \bibnamefont
			{Weis}}, \bibinfo {author} {\bibfnamefont {C.~C.}\ \bibnamefont {Lo}},
		\bibinfo {author} {\bibfnamefont {V.}~\bibnamefont {Lang}}, \bibinfo {author}
		{\bibfnamefont {A.~M.}\ \bibnamefont {Tyryshkin}}, \bibinfo {author}
		{\bibfnamefont {R.~E.}\ \bibnamefont {George}}, \bibinfo {author}
		{\bibfnamefont {K.~M.}\ \bibnamefont {Yu}}, \bibinfo {author} {\bibfnamefont
			{J.}~\bibnamefont {Bokor}}, \bibinfo {author} {\bibfnamefont {S.~A.}\
			\bibnamefont {Lyon}}, \bibinfo {author} {\bibfnamefont {J.~J.~L.}\
			\bibnamefont {Morton}}, \ and\ \bibinfo {author} {\bibfnamefont
			{T.}~\bibnamefont {Schenkel}},\ }\bibfield  {title} {\enquote {\bibinfo
			{title} {Electrical activation and electron spin resonance measurements of
				implanted bismuth in isotopically enriched silicon-28},}\ }\href {\doibase
		http://dx.doi.org/10.1063/1.4704561} {\bibfield  {journal} {\bibinfo
			{journal} {Appl. Phys. Lett.}\ }\textbf {\bibinfo {volume} {100}},\ \bibinfo
		{eid} {172104} (\bibinfo {year} {2012})}\BibitemShut {NoStop}%
	\bibitem [{\citenamefont {Feher}(1959)}]{Feher1959}%
	\BibitemOpen
	\bibfield  {author} {\bibinfo {author} {\bibfnamefont {G.}~\bibnamefont
			{Feher}},\ }\bibfield  {title} {\enquote {\bibinfo {title} {Electron spin
				resonance experiments on donors in silicon. i. electronic structure of donors
				by the electron nuclear double resonance technique},}\ }\href {\doibase
		10.1103/PhysRev.114.1219} {\bibfield  {journal} {\bibinfo  {journal} {Phys.
				Rev.}\ }\textbf {\bibinfo {volume} {114}},\ \bibinfo {pages} {1219--1244}
		(\bibinfo {year} {1959})}\BibitemShut {NoStop}%
	\bibitem [{\citenamefont {Ning}, \citenamefont {Selesnick},\ and\ \citenamefont
		{Duval}(2014)}]{Ning2014}%
	\BibitemOpen
	\bibfield  {author} {\bibinfo {author} {\bibfnamefont {X.}~\bibnamefont
			{Ning}}, \bibinfo {author} {\bibfnamefont {I.~W.}\ \bibnamefont {Selesnick}},
		\ and\ \bibinfo {author} {\bibfnamefont {L.}~\bibnamefont {Duval}},\
	}\bibfield  {title} {\enquote {\bibinfo {title} {Chromatogram baseline
				estimation and denoising using sparsity (beads)},}\ }\href {\doibase
		https://doi.org/10.1016/j.chemolab.2014.09.014} {\bibfield  {journal}
		{\bibinfo  {journal} {Chemom. Intell. Lab. Syst.}\ }\textbf {\bibinfo
			{volume} {139}},\ \bibinfo {pages} {156--167} (\bibinfo {year}
		{2014})}\BibitemShut {NoStop}%
	\bibitem [{\citenamefont {Lepine}(1972)}]{Lepine1972}%
	\BibitemOpen
	\bibfield  {author} {\bibinfo {author} {\bibfnamefont {D.~J.}\ \bibnamefont
			{Lepine}},\ }\bibfield  {title} {\enquote {\bibinfo {title} {Spin-dependent
				recombination on silicon surface},}\ }\href {\doibase 10.1103/PhysRevB.6.436}
	{\bibfield  {journal} {\bibinfo  {journal} {Phys. Rev. B}\ }\textbf {\bibinfo
			{volume} {6}},\ \bibinfo {pages} {436--441} (\bibinfo {year}
		{1972})}\BibitemShut {NoStop}%
	\bibitem [{\citenamefont {Poindexter}\ \emph {et~al.}(1981)\citenamefont
		{Poindexter}, \citenamefont {Caplan}, \citenamefont {Deal},\ and\
		\citenamefont {Razouk}}]{Poindexter1981}%
	\BibitemOpen
	\bibfield  {author} {\bibinfo {author} {\bibfnamefont {E.~H.}\ \bibnamefont
			{Poindexter}}, \bibinfo {author} {\bibfnamefont {P.~J.}\ \bibnamefont
			{Caplan}}, \bibinfo {author} {\bibfnamefont {B.~E.}\ \bibnamefont {Deal}}, \
		and\ \bibinfo {author} {\bibfnamefont {R.~R.}\ \bibnamefont {Razouk}},\
	}\bibfield  {title} {\enquote {\bibinfo {title} {{Interface states and
					electron spin resonance centers in thermally oxidized (111) and (100) silicon
					wafers}},}\ }\href {\doibase 10.1063/1.328771} {\bibfield  {journal}
		{\bibinfo  {journal} {J. Appl. Phys.}\ }\textbf {\bibinfo {volume} {52}},\
		\bibinfo {pages} {879--884} (\bibinfo {year} {1981})}\BibitemShut {NoStop}%
	\bibitem [{\citenamefont {Woodbury}\ and\ \citenamefont
		{Ludwig}(1960)}]{Woodbury1960}%
	\BibitemOpen
	\bibfield  {author} {\bibinfo {author} {\bibfnamefont {H.~H.}\ \bibnamefont
			{Woodbury}}\ and\ \bibinfo {author} {\bibfnamefont {G.~W.}\ \bibnamefont
			{Ludwig}},\ }\bibfield  {title} {\enquote {\bibinfo {title} {Spin resonance
				of transition metals in silicon},}\ }\href {\doibase 10.1103/PhysRev.117.102}
	{\bibfield  {journal} {\bibinfo  {journal} {Phys. Rev.}\ }\textbf {\bibinfo
			{volume} {117}},\ \bibinfo {pages} {102--108} (\bibinfo {year}
		{1960})}\BibitemShut {NoStop}%
	\bibitem [{\citenamefont {Marcos}\ \emph {et~al.}(2010)\citenamefont {Marcos},
		\citenamefont {Wubs}, \citenamefont {Taylor}, \citenamefont {Aguado},
		\citenamefont {Lukin},\ and\ \citenamefont {Sørensen}}]{Marcos2010}%
	\BibitemOpen
	\bibfield  {author} {\bibinfo {author} {\bibfnamefont {D.}~\bibnamefont
			{Marcos}}, \bibinfo {author} {\bibfnamefont {M.}~\bibnamefont {Wubs}},
		\bibinfo {author} {\bibfnamefont {J.~M.}\ \bibnamefont {Taylor}}, \bibinfo
		{author} {\bibfnamefont {R.}~\bibnamefont {Aguado}}, \bibinfo {author}
		{\bibfnamefont {M.~D.}\ \bibnamefont {Lukin}}, \ and\ \bibinfo {author}
		{\bibfnamefont {A.~S.}\ \bibnamefont {Sørensen}},\ }\bibfield  {title}
	{\enquote {\bibinfo {title} {Coupling nitrogen-vacancy centers in diamond to
				superconducting flux qubits},}\ }\href
	{http://link.aps.org/doi/10.1103/PhysRevLett.105.210501} {\bibfield
		{journal} {\bibinfo  {journal} {Phys. Rev. Lett.}\ }\textbf {\bibinfo
			{volume} {105}},\ \bibinfo {pages} {210501} (\bibinfo {year}
		{2010})}\BibitemShut {NoStop}%
	\bibitem [{\citenamefont {K\"opf}\ and\ \citenamefont
		{Lassmann}(1992)}]{Koepf1992}%
	\BibitemOpen
	\bibfield  {author} {\bibinfo {author} {\bibfnamefont {A.}~\bibnamefont
			{K\"opf}}\ and\ \bibinfo {author} {\bibfnamefont {K.}~\bibnamefont
			{Lassmann}},\ }\bibfield  {title} {\enquote {\bibinfo {title} {Linear {Stark}
				and nonlinear {Zeeman} coupling to the ground state of effective mass
				acceptors in silicon},}\ }\href {\doibase 10.1103/PhysRevLett.69.1580}
	{\bibfield  {journal} {\bibinfo  {journal} {Phys. Rev. Lett.}\ }\textbf
		{\bibinfo {volume} {69}},\ \bibinfo {pages} {1580--1583} (\bibinfo {year}
		{1992})}\BibitemShut {NoStop}%
\end{thebibliography}

%

\end{document}